\newcommand\Sul{Sul--Cu$_2$Cl$_4$}

\newcommand{\be}{\begin{equation}}
\newcommand{\ee}{\end{equation}}
\newcommand{\bea}{\begin{eqnarray}}
\newcommand{\eea}{\end{eqnarray}}

\newcommand{\p}{\partial}
\newcommand{\s}{\sigma}

\newcommand{\rd}{\mbox{d}}
\newcommand{\ri}{\mbox{i}}
\newcommand{\re}{\mbox{e}}

\documentclass[prb,twocolumn,floatfix,preprintnumbers,amsmath,amssymb,superscriptaddress]{revtex4}

\usepackage{graphicx}
\usepackage{dcolumn}
\usepackage{bm}

\begin{document}

\title{Excitations from a chiral magnetized state of a frustrated quantum spin liquid}

\author{A. Zheludev}
\altaffiliation{Present address: Laboratory for Neutron Scattering, ETH Z\"{u}rich \&
Paul Scherrer Institut, Villigen, Switzerland}
\affiliation{Neutron Scattering Sciences Division, Oak Ridge
National Laboratory, Oak Ridge, Tennessee 37831, USA.}
\author{V. O. Garlea}
\affiliation{Neutron Scattering Sciences Division, Oak Ridge
National Laboratory, Oak Ridge, Tennessee 37831, USA.}
\author{A. Tsvelik}
\affiliation{Department of Condensed Matter Physics and Materials Science, Brookhaven National Laboratory, Upton, NY 11973, USA.}
\author{L.-P.~Regnault}
\affiliation{CEA-Grenoble, INAC-SPSMS-MDN, 17 rue des Martyrs,
38054 Grenoble Cedex 9, France.}
\author{K. Habicht}
\affiliation{Helmholtz-Zentrum Berlin f\"{u}r Materialien und Energie, D-14109 Berlin, Germany.}
\author{K. Kiefer}
\affiliation{Helmholtz-Zentrum Berlin f\"{u}r Materialien und Energie, D-14109 Berlin, Germany.}
\author{B. Roessli}
\affiliation{Laboratory for Neutron Scattering, ETH Z\"{u}rich and
Paul Scherrer Institut, Villigen, Switzerland}
\date{\today}

\begin{abstract}
We study excitations in weakly interacting pairs of quantum spin
ladders coupled through geometrically frustrated bonds. The ground state is a
disordered spin liquid, that at high fields is replaced by an
ordered chiral helimagnetic phase. The spectra observed by high-field
inelastic neutron scattering experiments on the prototype compound
\Sul\ are qualitatively different from those in the previously studied
frustration-free spin liquids. Beyond the critical field
$H_c=3.7$~T, the soft mode that drives the quantum phase transition
spawns {\it two} separate excitations: a gapless Goldstone mode
and a massive magnon. Additional massive quasiparticles are
clearly visible below $H_c$, but are destroyed  in the ordered
phase. In their place one observes a sharply bound excitation continuum.
\end{abstract}

\pacs{75.10.Jm, 75.40.Gb, 73.43.Nq, 78.70.Nx}

\maketitle

\section{Introduction}


Gapped quantum-disordered antiferromagnets (AFs), also
known as ``spin liquids'', have become prototype materials for the
study of Bose-Einstein condensation and related phenomena.\cite{Giamarchi99,Rice,Giamarchi08, Matsumoto,IPA1,IPA2} By virtue of the Zeeman effect, the magnetic field directly affects the
chemical potential for the relevant lowest-energy triplet
excitations. The latter excitations can be viewed as bosons with hard-core
repulsion. Exotic quantum critical points are realized when the
gap energy for one member of the triplet is driven to zero at some
critical field $H_c$, and a macroscopic number of those magnons
get incorporated into the ground state. At $H>H_c$ one typically
observes a magnetically ordered state. However, due to the strong interactions the corresponding
excitation spectrum is nothing like that of conventional ordered
AFs.~\cite{IPA2}

Usually at $H>H_c$ the magnon branch that goes
soft at the quantum critical point is replaced by a gapless
collective Goldstone mode.\cite{Matsumoto,IPA1,IPA2} For all models and prototypical
materials studied to date, the two other members of the original
excitation triplet survive as sharp gap excitations. Thus, the
low-energy spectrum remains dominated by three long-lived
quasiparticles. In the present work we study the spin dynamics in
a gapped quantum AF with strong geometric frustration of magnetic
interactions. In a stark contrast to the behavior of
frustration-free spin liquids, we find that above $H_c$ the soft
mode gives rise to two distinct excitation branches, while the
higher-energy gapped magnons become unstable and are replaced by a
broad excitation continuum.

\begin{figure}[tbp]
\includegraphics[width=3.5in]{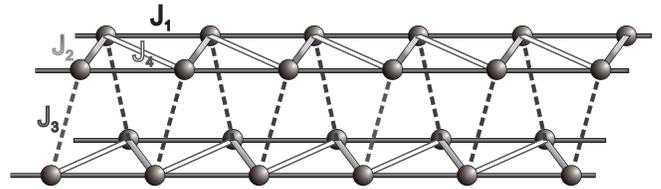}
\caption{A schematic representation of the magnetic interaction topology in \Sul. $J_1$ denotes the ``leg''  coupling, and $J_2$, $J_3$ and $J_4$ are the three distinct ``rung'' interactions.}
\label{struc}
\end{figure}

Our prototype material, \Sul,~\cite{Fujisawa,Garlea08,Garlea09} is a quasi-one-dimenisonal
Heisenberg AF with a singlet ground state and a spin  gap of
$\Delta_0=0.52$~meV. The corresponding network of $S=1/2$
Cu$^{2+}$ spins is best described as an array of 4-leg spin tubes with
dominant AF nearest-neighbor coupling along the legs and
several weaker rung interactions of comparable strength. The
tube's legs run along the crystallographic $c$ axis of the
triclinic $P\bar{1}$ structure. The corresponding interaction topology is
illustrated in Fig.~\ref{struc}. Each spin tube consists of two
spin ladders, $J_1$ and $J_3$ being the leg and the rung exchange
constants, respectively. The two ladders are coupled via
exchange constants $J_2$ and $J_4$, that are in obvious geometric
frustration with $J_1$. Even though the ground state is
disordered, frustration ensures that dynamic one-dimensional spin
correlations in \Sul\ are peaked at incommensurate wave vectors
$l_0 = 0.5 -\delta, 1.5 + \delta, \delta = 0.022(2)$.~\cite{Garlea08} The
dispersion of the gap excitation along the legs of the tubes can be approximated as
 \bea
 (\hbar \omega_\mathbf{q})^2 = \Delta_0^2 + v_0^2({\bf qc} - 2\pi l_0)^2,
 \label{spectrum}
 \eea
where $v_0 \sim 14$~meV is the spin wave velocity. Any dispersion
transverse to the $c$ axis is undetectably small. Previously we
have demonstarated that the application of a magnetic field
exceeding $H_c=3.7$~T destroys the spin liquid and induces an
ordered helimagnetic state with an incommensurate propagation
vector.~\cite{Garlea09} The incommensurability along the $c$ axis exactly
corresponds to the minimum of the dispersion at zero field. The
present work deals with the spin dynamics on either side of this
field-induced phase transition.

\section{Experimental procedures}
Inelastic neutron scattering measurements were performed on the
same fully deuterated single crystal samples as those used in
Ref.~\onlinecite{Garlea08}. Two series of experiments were carried out at
the V2-Flex and TASP cold neutron spectrometers at HMI and PSI,
respectively. In Setup 1 we used $E_f=3.7$~meV fixed final energy
neutrons, while the sample environment was a 14.5~T split-coil
cryomagnet with a dilution refrigerator insert. Data in Setup 2
were collected with $E_f=3$~meV and a similar 9~T setup. In both
cases we utilized pyrolitic graphite (PG) vertically focused
monochromators and horizontally focused analyzers, as well as Be
filters positioned after the sample. The field was in all cases
applied along the $b$ axis and scattering data were collected in
the $(h,0,l)$ reciprocal-space plane. Most of the data were measured in
constant-$q$ scans at the wave vector transfers where the
previously measured 1D magnon dispersion is a minimum:
$q$=($h,0,0.48$). Momentum transfers along the $a^\ast$ axis were chosen to
optimize wave vector resolution along the $c^\ast$ direction. The
background was measured at the wave vectors ($h,0,0.6$), where no
magnetic contribution is expected, due to a very steep dispersion
along the crystallographic $c^\ast$ direction. These background
scans were fit to a constant plus a Gaussian function centered at
zero energy transfer, to account for fast neutron background and
elastic incoherent scattering from the sample and sample
environment, respectively. The resulting fitted background
function was subtracted from the signal scans.

\section{Experimental results}

\begin{figure}[tbp]
\includegraphics[width=3.3in]{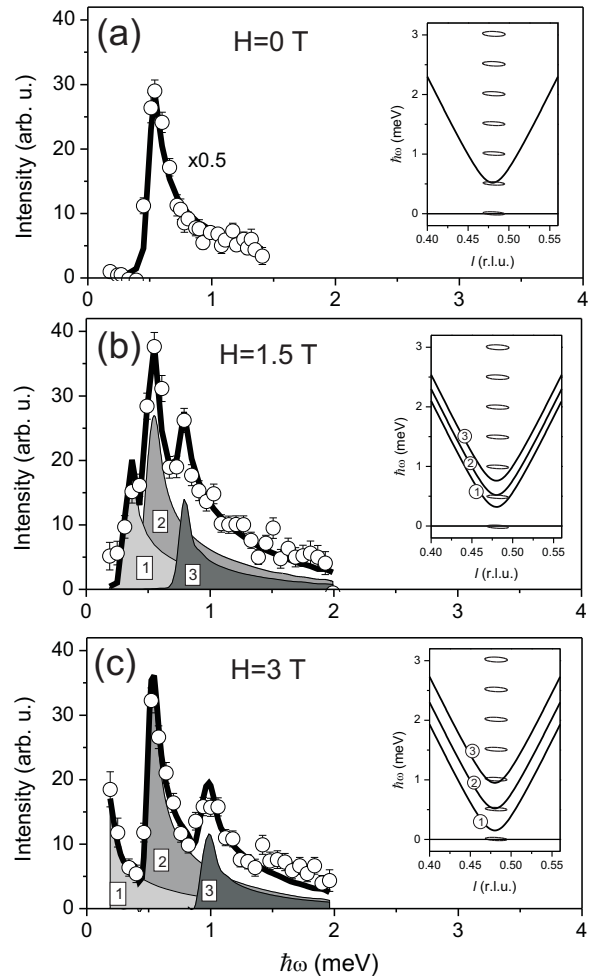}
\caption{Field dependence of background-subtracted inelastic scattering at the 1D AF zone-center (0.5, 0, 0.48), collected at $H<H_c$ and T $\approx$ 70~mK. Solid lines are model cross section fits, as described in Section~\ref{modeling}. Shaded areas are partial contributions of three separate excitation branches. Inserts: Dispersions of the three field-driven excitation branches in \Sul\, and evolution of the calculated FWHM resolution ellipsoids plotted in projection onto the ($l$, $\hbar\omega$) plane.}
\label{INSdata1}

\end{figure}

Typical background-subtracted scans collected at $H<H_c$ are shown
in Fig.~\ref{INSdata1}. As the external magnetic field is
increased, the single peak seen at $H=0$ (Ref.~\onlinecite{Garlea08},
Fig.~3) splits into three components. The gap energy of the
central and most intense component (mode 2) is field independent.
The energies of modes 1(3) decrease (increase) with the field,
 as expected for Zeeman splitting of the  $S=1$
excitation triplet. The gap in the lower branch (mode 1)
 approaches  zero energy as $H\rightarrow H_c$.

\begin{figure}[tbp]
\includegraphics[width=3.3in]{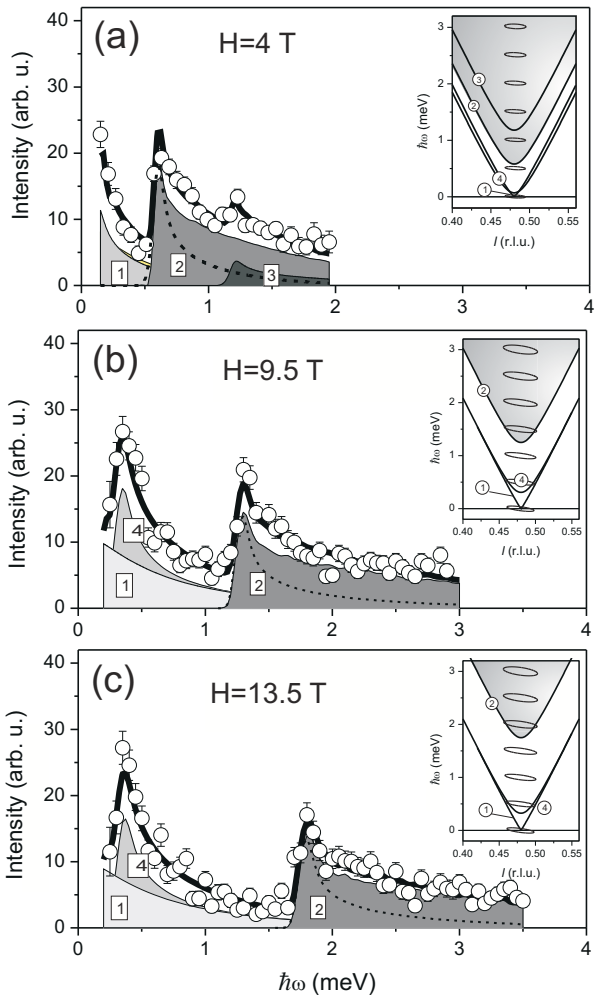}
\caption{Background-subtracted neutron spectra measured in \Sul\ at the 1D AF zone-center (0.5, 0, 0.48) in
various applied magnetic fields above the critical field, $H>H_c$, and 150~mK. The lines, shaded areas, and graphical representations of the resolution functions are as in Fig.~\ref{INSdata1}. The shaded area in the inserts indicates a continuum scattering, with the corresponding solid line depicting the lower bound of the continua.}
\label{INSdata2}
\end{figure}

For magnetic fields exceeding $H_c=3.7$~T (Fig.~\ref{INSdata2}) the measured
excitation spectrum undergoes some drastic changes. As can be seen
in Fig.~\ref{INSdata2}a, just above the critical field, at $H=4$~T,
mode 3 becomes very weak and is, in fact, barely visible. It is
totally absent for all other values of magnetic field applied in
our experiments (5~T, 8~T, 9.5~T and 13.5~T). The gap energy for
mode 2 starts to increase with increasing $H$. Finally, mode 1,
being critical at $H=H_c$, re-acquires an energy gap gap at
higher fields (Fig.~\ref{INSdata2}~b,c). This gap, however, increases
rather slowly with the field not exceeding  0.3~meV even at the
highest attainable field of 13.5~T.

\section{Theoretical considerations}
The re-opening of the spin gap at $H>H_c$ is, at the first glance,
puzzling. Indeed, in those spin liquids that undergo similar
field-induced ordering transitions the gap re-appears in the high
field phase due to magnetic anisotropy effects. In that case the
transition is of the Ising, rather than the BEC, universality class.
However, in the case of \Sul, magnetic anistropy is negilgibly
small, as discussed in Ref.~\onlinecite{Garlea09}. Moreover, since
the helimagnetic ordering occurs at an {\it incommensurate} wave
vector, one can rigorously prove that there {\it must} be a
gapless ``sliding mode'' (``phason''). This remains true
even if anisotropy is present. The sinusoidal spin spiral that is
the ground state in the isotropic case, will be distorted by
the anisotropy to form a soliton lattice. That soliton array, however,
being incommensurate, will retain the ability to slide freely
relative to the underlying crystal structure.

To fully understand the nature of the spin gap in \Sul, we shall
develop a field-theoretical description of the low-energy spin
correlations in a pair of magnetized spin ladders coupled with
geometrically frustrated interactions.

\subsection{Inter-ladder interactions}
We start with writing down the spin Hamiltonian for the frustrated
spin tube geometry shown in Fig.~\ref{struc}

 \bea
 H & =  & \sum_{n; a=1,2}\left[J_1{\bf S}^a_n{\bf S}^a_{n+1} + J_1{\bf T}^a_n{\bf T}^a_{n+1}+  J_3{\bf S}_n^a{\bf T}_n^a \right] + \nonumber\\
 & + &\sum_n \left[  J_4{\bf S}_n^1{\bf S}_n^2 + J_2{\bf S}_{n+1}^1{\bf S}_n^2 +\right.\nonumber \\
 & & +  \left. J_2{\bf T}_n^1{\bf T}_n^2 + J_4{\bf T}_{n+1}^1{\bf T}_n^2 \right]
 \label{lat}
 \eea

Here $a=1,2$ labels the two structural ladders, while
$\mathbf{S}_n$ an $\mathbf{T}_n$ are spin operators on the
corresponding legs. We direct the reader's attention to the fact
that the ratio of spin gap $\Delta_0$ to the magnetic excitations
bandwidth $v_0$ in \Sul\ is much smaller than in most known spin
ladders. This is likely to be a consequence of geometric
frustration. Whatever the cause, it makes this particular system
ideal for application of the field theory methods developed in Ref.
\onlinecite{Shelton}.

We shall treat Hamiltonian (\ref{lat}) as one describing two
weakly interacting ladders. Assuming $J_1\gg J_2, J_3, J_4$, there
are two cases to consider: (i) $J_3 \gg |J_2-J_4|\equiv J_\bot$.
Here one ladder is composed of spins
${\bf S}_{1,2}$ and the other of spins ${\bf T}_{1,2}$. Such ladders have spectral gaps
$\Delta \sim J_3$. The second sum in (\ref{lat}) represents a frustrated interaction and in the gapped
phase can be considered as a perturbation.   This is justified by
the fact that the term with $J_3$ has smaller scaling dimension than
$J_2,J_4$  hence generating  a stronger coupling. (ii) Alternatively, one may consider
$|J_2 - J_4| \gg J_3\equiv J_\bot$. Then the two ladders are
composed of ${\bf S}$ and ${\bf T}$ spins, respectively. The spin
gap is of the order of $|J_2 -J_4|$. In either scenario, the
inter-ladder interactions $J_\bot$ are irrelevant at small
magnetic fields, yet  become relevant when the external field
drives the single-ladder gap to zero.

To obtain the continuum limit of the lattice model (\ref{lat}), we
approximate the spin operators as follows:
\bea
&& {\bf S}_n^a \approx {\bf M}^a(x) + (-1)^n{\bf n}^a(x), ~~ a=1,2\nonumber\\
&& {\bf T}_n^a \approx {\bf L}^a(x) + (-1)^n{\bf l}^a(x), ~~ a=1,2
\label{defs}
\eea
where ${\bf l}, {\bf n}$  and ${\bf L}, {\bf M}$
are operators with scaling dimensions 1/2 and 1 respectively. It
turns out that only antibonding combinations of the staggered
magnetizations become critical in high magnetic fields. Therefore
only such combinations will generate relevant inter-ladder
coupling at $H > H_c$. They are defined as ${\bf N}_a = {\bf n}^a
- {\bf l}^a$ ($a=1,2$) and ${\bf N}_1 = {\bf n}_1 - {\bf n}_2,  {\bf N}_2 =
{\bf l}_1 - {\bf l}_2$, for cases (i) and (ii), respectively.

Substituting these definitions into (\ref{lat}) and keeping only
the antibonding staggered magnetization terms, we obtain the
following inter-ladder interaction:
 \bea V = \gamma \int \rd x
 \left[{\bf N}_1\p_x{\bf N}_2 - {\bf N}_2\p_x{\bf N}_1\right],
 \label{V}
 \eea
 where $\gamma \sim J_\bot/J_1$.

\subsection{Ginsburg-Landau model}
Interaction (\ref{V}) are easier to treat in the spin-liquid phase
when the spectrum is gapped. Here we shall follow Affleck
\cite{Affleck} who suggested to  describe the low lying modes of
spin ladder by an effective Ginzburg-Landau model with the
Lagrangian density
\bea
{\cal L} = \frac{1}{2}(\p_{\tau}{\mathbf\Phi})^2 +
\frac{\tilde{v}^2}{2}(\p_x{\mathbf\Phi})^2 +
\frac{\tilde\Delta^2}{2}{\mathbf\Phi}^2 + g({\mathbf\Phi}^2)^2
\eea
where ${\mathbf{\Phi}} \sim {\bf N}$, and the bare parameters
$\tilde{v}$, $\tilde\Delta$  are of the same order as the velocity
and gap in the actual single-ladder spectrum. It is
straightforward to show that the introduction of the interaction
(\ref{V}) generates two modes with the spectrum
 \bea
 E^2_{\pm} = \Delta^2 +
 v^2({\bf qc} - \pi \pm \delta)^2
 \label{spectrum1}
 \eea
where $\Delta, v$ are the renormalized gap energy and spin wave
velocity, respectively, and $\delta \sim \gamma$. The
incommensurate gapped dispersion is fully in agreement with the
existing data on \Sul.

When the magnetic field exceeds the spin gap the transverse
components of the staggered magnetizations become critical.
Interaction (\ref{V}) becomes relevant, but its form is highly
non-standard from the field theory point of view since the
corresponding interaction density  has nonzero conformal spin
(that is its left and right conformal dimensions do not coincide).
Attempts to treat such interactions using standard field theory
methods have met with mixed success \cite{Essler}. Below we shall
adopt a different approach. Again, following Affleck
\cite{Affleck}, close to the critical field we will describe each
spin ladder using the Landau-Ginzburg theory with the $|\phi|^4$
interaction. When applying this formalism to our case, one replaces
$N^z = (1- \psi^+\psi), N^{\pm} = (\psi, \psi^+)$, where $\psi$
are bosonic fields  and writes the Lagrangian as
 \bea
 && {\cal L} = \psi^+_a(\p_{\tau} - \mu)\psi_a + \frac{1}{2m}\p_x\psi^+_a\p_x\psi_a + g(|\psi_1|^4 + |\psi_2|^4) + \nonumber\\
 && J_{\perp}(\psi_1^+\p_x\psi_2 - \psi_2^+\p_x\psi_1),
 \eea
where  $\mu \sim H -H_c$, $m \sim \Delta$, and the coupling
 $g$ is assumed to be large.

\subsection{Incommensurate order and spectrum at $H>H_c$}

We now do the substitution
 \bea \psi =
 \frac{1}{\sqrt 2}(I + \ri\s_x)\tilde\psi \eea and \bea
 \tilde\psi_{1,2} = \re^{\pm \ri Qx}\eta_{1,2},
 \eea
where $\s^x$ is the Pauli matrix and $Q \sim  \gamma$. Then the interaction density becomes
 \bea
 && g(|\psi_1|^4 + |\psi_2|^4) \rightarrow \nonumber\\
 & \rightarrow & \frac{g}{2}\left[(|\eta_1|^2 + |\eta_2|^2)^2 - (\eta^+_1\eta_2\re^{2\ri Qx} - \eta^+_1\eta_2\re^{-2\ri Qx})^2\right] \rightarrow \nonumber\\
 & \rightarrow & \frac{g}{2}\left[(|\eta_1|^2 + |\eta_2|^2)^2 + 2|\eta_1|^2|\eta_2|^2\right],
 \eea since at large $Q$ the
oscillatory terms are wiped out. when one integrates over $x$.  In terms
of $\eta_a$'s the ground state of the resulting model is a ferromagnet. In other words,
only one species of bosons condenses, {\it e. g.}, $\eta_1$. The
Hamiltonian for this field is just the bosonic one with a
point-like repulsion. So the excitation spectrum at small wave
vectors is linear in wave vector corresponding to the phase fluctuations of the condensate.
Bosons of the second flavor remain massive with
the energy gap $\Delta_2\sim |\eta_1|^2\sim \mu$.

Now we have to go back to the original variables. It turns out that the magnetic structure observed experimentally in
\Sul\ corresponds to the case denoted above as (i).
For that case we
get the following picture of the soft spin modes:
 \bea
 && {\bf n}_1^{\pm} \approx  - {\bf l}_1^{\pm} \approx {\bf N}_1^{\pm} = (\eta_1\re^{\ri Qx}, \eta_1^*\re^{-\ri Qx})\\
 && {\bf n}_2^{\pm} \approx  - {\bf l}_2^{\pm} \approx {\bf N}_2^{\pm} =
 \ri(\eta_1\re^{\ri Qx}, -\eta_1^*\re^{-\ri Qx}).
 \eea
All spin fluctuations occur in the plane perpendicular to the
magnetic field. Along the $J_3$ bonds the spins are antiparallel.
Spins along $J_2$ or $J_4$ are perpendicular to each other. The
propagation vector along the leg axis is $2\pi l_0 = \pi - Q$, so
$l_0$ is close to 1/2. Similarly, for (ii):
 \bea
 && n_1^{\pm} \approx  - n_2^{\pm} \approx N_1^{\pm} = (\eta_1\re^{\ri Qx}, \eta_1^*\re^{-\ri Qx})\\
 && l_1^{\pm} \approx  - l_2^{\pm} \approx N_2^{\pm} =
 \ri(\eta_1\re^{\ri Qx}, -\eta_1^*\re^{-\ri Qx}).
 \eea
here the spins are parallel along $J_2$, and perpendicular on the
$J_3$ bond.

The model thus qualitatively reproduces the diffraction result for the elastic modes,
but also predicts that at $H>H_c$ the soft mode splits into {\it
two} components. One, as required, is gapless. The other mode has
a gap that scales as the square of the transverse ordered moment.
Below we shall show that this prediction is indeed fully
consistent with the inelastic data.

\section{Data analysis}\label{modeling}
A quantitative analysis of the inelastic neutron data involves
fitting the measured scans to a parameterized model cross section
function, numerically convoluted with the 4-dimensional resolution
function of the spectrometer. We start with the simpler regime
$H<H_c$.

\subsection{Low fields} For the vicinity of the minima of the one-dimensional
dispersion where all scans were collected, the cross section
was written in the single-mode approximation assuming three
separate excitation branches corresponding to $S_z=0,\pm 1$. Since
the Zeeman term commutes with the Heisenberg Hamiltonian, the
spectrum at any $H<H_c$ will be exactly as at $H=0$, except for a
constant shift in the energies of the $S_z=1$ and $S_z=-1$
magnons. With this in mind, the dispersion relations for mode
$\alpha$ ($\alpha=1,2,3$) was written as:
\begin{equation}
 \hbar \omega_{\alpha,\mathbf{q}}  =  \hbar
 \omega_{\mathbf{q}}+\Delta_\alpha-\Delta_0. \label{disp}
 \end{equation}
The spin wave velocity and zero-field gap were fixed at
$v_0=14$~meV and $\Delta_0=0.52$~meV, respectively, as previously
measured in zero field. The gap energies $\Delta_\alpha$ for each
mode were assumed to be field-dependent. The single-mode cross
section was then written as:
\begin{equation}
 \frac{d^2 \sigma}{d \Omega d E'} \propto \sum_\alpha |f(\mathbf{q})|^2
 \frac{A_\alpha}{ \hbar
 \omega_{\mathbf{q}} } \delta(\omega-\omega_{\alpha,\mathbf{q}}). \label{sqw1}
\end{equation}
Here are $A_\alpha$ are separate intensity prefactors for each of
the three modes. This model, when convoluted with the known
instrument resolution, fits all the scans collected at $H<H_c$
rather well. Such fits are shown in heavy solid lines in
Fig.~\ref{INSdata1}. Contributions due to the three components of
the spectrum are represented by shaded areas. The asymmetric peak
shapes with extended ``tails'' on the high-energy side are
entirely due to resolution effects.

\subsection{$H>H_c$: low-energy modes} Finding an appropriate dispersion
relation for the two theoretically predicted descendants of the
soft mode in the regime $H>H_c$ is not straightforward. The
guidance provided by the calculations in the previous section is
limited, as they are only applicable in the direct proximity of
the ordering vector.

For the gapless mode it is clear that at $H=H_c$ one should still
be able to employ Eqs.~\ref{disp} and \ref{spectrum} with
$\Delta_1=0$. However, for $H>H_c$ the dispersion near the
ordering vector should be linear, with the slope progressively
increasing with $H-H_c$. In the strongly 1D case like that of
\Sul, one can expect the dispersion relation to follow the lower
bound of continuum excitations in {\it isolated} ladders. The
latter were investigated in Ref.~\onlinecite{Sushkov} and can be
represented by the same parabolic curve as at $H=0$, but offset
both in energy and momentum. The ``gap'' $\Delta_1$ becomes
negative. The corresponding dispersion and neutron cross section
are still given by Eqs.~\ref{disp} and \ref{sqw1}, respectively,
but the definition of $\hbar \omega_{\mathbf{q}}$ in (\ref{disp})
is changed to:

 \begin{eqnarray}
 (\hbar \omega_{\mathbf{q}})^2 &=& \Delta_0^2 +  v_0^2\left(|{\mathbf{q}}{\mathbf{c}}-2\pi l_0|+\kappa \right)^2,\label{disp2}\\
 \kappa^2 & = &  \frac{\Delta_0
 |\Delta_1|}{v_0^2}\left(2-\frac{\Delta_1}{\Delta_0}\right).\\
 \nonumber
\end{eqnarray}
Typical dispersion relations calculated using this equation are
the lowest-energy curves plotted in the insets in
Fig.~\ref{INSdata2}.

We shall label the theoretically predicted gapped descendent of
the soft mode with the index $\alpha=4$. It appears reasonable to
approximate its dispersion with Eqs.~\ref{spectrum} and
\ref{disp}, as at $H<0$. The gap $\Delta_4$ can be accurately
determined from our scans that show a peak at the corresponding
energy. In contrast, the parameter $\Delta_1$ actually represents
the velocity at $\mathbf{q}\mathbf{c}\rightarrow l_0$. It can not
be accurately measured due to limitations imposed by the
experimental resolution. On the other hand, it can not be totally
disregarded either, as it affects the observable high-energy
resolution ``tail'' of the gapless mode. In order to avoid dealing
with an over-parameterized model, a compromise was reached by
somewhat arbitrarily postulating $\Delta_4=|\Delta_1|$. Indeed,
both parameters are zero at $H=H_c$ and are expected to increase
in absolute values at higher fields. As represented by the heavy
solid lines in the lower-energy range of the plots shown in
Fig.~\ref{INSdata2}, this model, though somewhat artificial,
reproduces the experimental data remarkably well. In the same
figure the light grey shaded areas are partial contributions of
the gapless and gapped components (modes 1 and 4), respectively.

\subsection{$H>H_c$: high energies}
Upon crossing the critical field, the central component of the
excitation triplet undergoes some drastic changes. This mode is
the strongest and sharpest feature of the spectrum at $H<H_c$
(Fig.~\ref{INSdata1}c), but is replaced by a much broader peak above the
transition (Fig.~\ref{INSdata2}a), measured using in the same experimental
configuration). The single-mode approximation (\ref{sqw1}) gives
excellent fits to the data at low fields, but totally fails to
describe the shape of the middle mode at $H>H_c$ (Fig.~\ref{INSdata2}, dashed
curves). This leads us to the conclusion that the quasiparticle
description of this part of the spectrum breaks down in the
high-field phase. Instead, the excitations are a broad continuum
of states. By analogy with the M\"{u}ller ansatz that describes
continuum excitations in gapless spin chains, we used the
following empirical expression to approximate the corresponding
contribution ($\alpha=2$) of this spectral feature to the neutron
cross section:
\begin{equation}
 \frac{d^2 \sigma}{d \Omega d E'} \propto A_2 |f(\mathbf{q})|^2
(\omega-\omega_{2,\mathbf{q}})^{-\zeta}\Theta(\omega-\omega_{2,\mathbf{q}}).\label{cont}
\end{equation}
In this formula $\hbar \omega_{2,\mathbf{q}}$ is still given by
Eq.~\ref{disp}, and describes the lower bound of the continuum. A
new parameteter $\zeta$ characterizes how steeply the intensity
falls off with energy. With the value $\zeta=0.30(6)$ refined at
$H=9.5$~T, this formula yields much better fits to the data at all
fields at $H>H_c$ (Fig.~\ref{INSdata2}, solid curves) than the single-mode
approximation (dashed curves).

The highest-energy component of the original triplet of gap
excitations is also strongly affected by the transition. As
clearly seen in Fig.~\ref{INSdata2}, its intensity drops dramatically and
it is no longer observed at $H>4$~T. To describe the mode at that
field however, in our analysis we employed an expression similar
to that for the middle mode, the parameter $\zeta$ fixed at the
same value.

\subsection{Fits to experimental data}
\begin{figure}[tbp]
\includegraphics[width=3.3in]{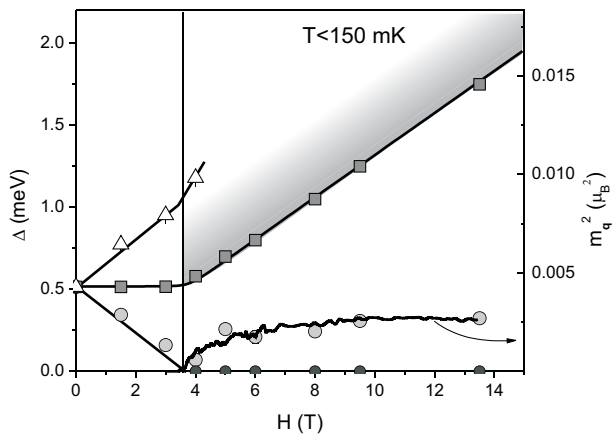}
\caption{Field dependence of the excitation energies at the 1D AF
zone center in \Sul. The points are the data from the INS experiments and the solid lines are guides for the eye. At $H > H_c$, the central excitation branch is replaced by a sharp continuum represented by the  shaded band. The field dependence of the new gapped excitation (above 3.7~T) is compared with that of the square of the transverse ordered moment, from Ref.~\onlinecite{Garlea09}.}
\label{gapvsH}
\end{figure}

The model cross sections described above were numerically
convoluted withe the spectrometer resolution function, calculated
in the Popovici approximation.\cite{Popovici} They were then used in the
least-squares analysis of constant-$Q$ scans collected at each
value of the applied field. The resulting fits are shown as the  solid
lines in Figs.~\ref{INSdata1} and ~\ref{INSdata2}. Partial contributions of each
spectral component are represented by the shaded areas. The insets
show the dispersion relations of all components calculated using
the parameters determined in the fitting process. The shaded area
signifies continuum scattering, with the corresponding solid line
indicating the lower bound. In the same plot we show the evolution
of the experimental resolution ellipse in the course of each
constant-$Q$ scan. The gap energies obtained in these fits are
plotted vs. applied magnetic field on Fig.~\ref{gapvsH} (symbols).

\section{Discussion}
Despite fact that our data analysis is based on several somewhat
arbitrary assumptions regarding the form of the dynamic structure
factor, the extracted field dependencies of the gap energies should
be quite reliable. Indeed, the gaps are robustly identified by a
sharp jump in the neutron intensity that occurs in the energy window
defined by the resolution function. The fact that the central
excitation branch is replaced by a sharp continuum at $H>H_c$ is
also a solid experimental finding, as is the disappearance of the
higher-energy mode.

The linear Zeeman splitting of the triplet at $H<H_c$ occurs
exactly as expected: the system is not ordered and the Zeeman term
commutes with the Heisenberg Hamiltonian. For $H>H_c$, the central
prediction of our Ginsburg-Landau analysis is that the energy of
the new gapped excitation scales as the square of the transverse
ordered moment. This is indeed consistent with the experiment. In
Fig.~\ref{gapvsH} the heavy solid line, and the corresponding
right-hand-side Y axis, represent the field dependence of the
latter, as previously measured experimentally.\cite{Garlea09} To within a
constant scaling factor, this curve exactly reproduces the field
dependence of the re-opening gap $\Delta_4(H)$ measured in this
work.

The dramatic broadening of the higher-energy excitations at $H>H_c$ is not captured by our hydrodynamic model. While a further theoretical study is clearly required, we note that the phenomenon resembles the one  previously observed in the bond-alternating S = 1 chain system NTENP.\cite{ntenp} In the latter system it was attributed to new magnon decay channels that are opened by the lifting of rotational symmetry at $H_c$. A similar mechanism may at work  in \Sul.

\section{Conclusion}
In summary, we suggest that  the geometric frustration and the coupled-ladder topology
of \Sul\ are responsible for a complex and very unusual excitation
spectrum in the magnetized state. At the same time, triplet
low-energy excitations at $H<H_c$ are rather typical for an
isotropic spin liquid.

\section{Acknowledgements}

The work at Oak Ridge National Laboratory was sponsored by the Scientific User Facilities Division, Office of Basic Energy Sciences, U. S. Department of Energy. U.S. DOE. ORNL is operated by UT-Battelle, LLC for the U.S. DOE under Contract No. DE-AC05-00OR22725. AMT acknowledges the support from US DOE under contract number DE-AC02 -98 CH 10886.


\begin{thebibliography}{}
\bibitem{Giamarchi99} T. Giamarchi and A. M. Tsvelik, Phys. Rev. B {\bf 59}, 11398 (1999).
\bibitem{Rice} M. Rice, Science {\bf 298}, 760 (2002).
\bibitem{Giamarchi08} T. Giamarchi, C. Ruegg, O. Tchernyshyov, Nature Physics, {\bf 4}, 198 (2008).
\bibitem{Matsumoto} M. Matsumoto, B. Normand, T. M. Rice, and M. Sigrist, Phys. Rev. Lett. {\bf 89}, 077203 (2002).
\bibitem{IPA1} V. O. Garlea, A. Zheludev, T. Masuda, H. Manaka, L.-P. Regnault, E. Ressouche, B. Grenier, J.-H. Chung, Y. Qiu, K. Habicht, K. Kiefer and M. Boehm, Phys. Rev. Lett. {\bf 98}, 167202 (2007).
\bibitem{IPA2}  A. Zheludev, V. O. Garlea, T. Masuda, H. Manaka, L.-P. Regnault, E. Ressouche, B. Grenier, J.-H. Chung, Y. Qiu, K. Habicht, K. Kiefer and M. Boehm, Phys. Rev. B {\bf 76}, 054450 (2007).
\bibitem{Fujisawa} M. Fujisawa, J.-I. Yamaura, H. Tanaka, H. Kageyama, Y. Narumi, and K. Kindo, J. Phys. Soc. Jpn. {\bf 72}, 694 (2003).
\bibitem{Garlea08} V. O. Garlea, A. Zheludev, L.-P. Regnault, J.-H. Chung, Y. Qiu, M. Boehm, K. Habicht, M. Meissner, Phys. Rev. Lett., {\bf 100}, 037206 (2008).
\bibitem{Garlea09} V. O. Garlea, A. Zheludev, K. Habicht, M. Meissner, B. Grenier, L.-P. Regnault, and E. Ressouche, Phys. Rev. B, {\bf 79}, 060404(R) (2009).
\bibitem{Shelton} D. G. Shelton,  A. A. Nersesyan and A. M. Tsvelik, Phys. Rev. B {\bf 53}, 8521 (1996).
\bibitem{Affleck} I. Affleck, Phys. Rev. B {\bf 41}, 6697 (1990).
\bibitem{Essler} A. A. Nersesyan, A. O. Gogolin, F. H. L. Essler, Phys. Rev. Lett. {\bf 81}, 910 (1998).
\bibitem{Sushkov} O. P. Sushkov and V. N. Kotov, Phys. Rev. Lett. {\bf 81}, 1941 (1998).
\bibitem{Popovici} M. Popovici, Acta Crystallogr. {\bf A31}, 507 (1975).
\bibitem{ntenp} L.-P. Regnault, A. Zheludev, M. Hagiwara, A. Stunault, Phys. Rev. B, {\bf 73}, 174431 (2006).

\end{thebibliography}
\end{document}